\documentstyle[prd,preprint,aps,epsfig,floats]{revtex}
\draft
\newcommand{\beq}{\begin{equation}}
\newcommand{\eeq}{\end{equation}}
\newcommand{\bea}{\begin{eqnarray}}
\newcommand{\eea}{\end{eqnarray}}
\begin{document}

\title{Study of Nondiagonal Parton Distribution Models} 

\author{Andreas Freund, Vadim Guzey}

\address{
Department of Physics,  The Pennsylvania State University\\
University Park, PA  16802, U.S.A.}

\maketitle
 
\begin{abstract}
In this paper we examine predictions from different models of nondiagonal 
parton distributions. This will be achieved by examining whether certain 
predictions of relationships between diagonal and nondiagonal parton 
distributions also hold after having evolved the different distributions.    
\end{abstract}

\section{Introduction}
\label{intro}

In recent years
 deeply virtual Compton scattering (DVCS), hard exclusive
electroproduction processes and the distributions which contain the 
nonperturbative information for those processes, namely the nondiagonal or
nonforward parton distributions, together with their evolution have been a 
target of intense studies 
\cite{1,2,3,4,5,6,6a,7,8,9,10,11,12,12a,13,14,15,16,17,18,19}.

Nondiagonal parton distributions, in particular, have attracted a lot of 
attention and numerical studies as to their behavior under their
leading order (LO) evolution 
were conducted recently \cite{5,6,6a}.  Also recently, the 
next-to-leading order (NLO) 
generalized Efremov-Radyushkin-Brodsky-Lepage anomalous dimensions were computed
employing conformal 
symmetry arguments \cite{12} making NLO studies in the near future possible. 
The NLO kernels for the DGLAP region ($x_1>\Delta$) of the 
nondiagonal parton distributions can be reconstructed via the method described
in Ref.\ \cite{11a}.

In the above mentioned numerical studies, a particular ansatz for the initial 
distribution for nondiagonal evolution was made, namely that the 
nondiagonal and diagonal initial distributions were taken to be equal in the 
normalization point. As was pointed out in \cite{20} this ansatz cannot be 
justified except at very low normalization point $Q_0$ where the parton 
distributions are essentially flat. In the following, we will explore a 
different ansatz and investigate whether predictions based on this ansatz hold
after evolution. It is crucial for the predictive power of theoretical 
calculations of exclusive, hard diffractive processes to have sensible 
nonperturbative models of nondiagonal parton distributions as long as they 
have not been extracted from experiment.    

The paper is structured the following way. In  Sec.\ \ref{sec:ba} we recap
a few basic definitions about nondiagonal parton distributions and the involved
kinematics, in Sec.\ \ref{sec:mod} we will briefly explain the model under 
consideration as derived by Radyushkin \cite{20} and the predictions about the
relationship between nondiagonal and diagonal parton distributions, in 
Sec.\ \ref{sec:res} we will give the results of our study, making some remarks 
about the behavior of nondiagonal parton distributions in NLO and, finally, in 
Sec.\ \ref{sec:concl} we give concluding remarks.

\section{Basics}
\label{sec:ba}

The kinematics as encountered in the appearance of nondiagonal parton 
distributions can be most easily described by the DVCS process 
$\gamma^*(q) + P(p) \rightarrow \gamma(q') + P(p')$.
The fact that the initial and final state protons have different momenta 
leads to the introduction of nondiagonal parton distributions. The nondiagonal
character of these distributions can be seen in the definitions of 
nondiagonal parton distributions as Fourier transforms of matrix elements of 
bilocal, renormalized light-cone operators (see Ref.\ \cite{2,3,4} and below)
, the light-cone operators being sandwiched between states of different 
momenta 
as compared to the diagonal case where the states have the same momentum.

\begin{figure}
\centering
\mbox{\epsfig{file=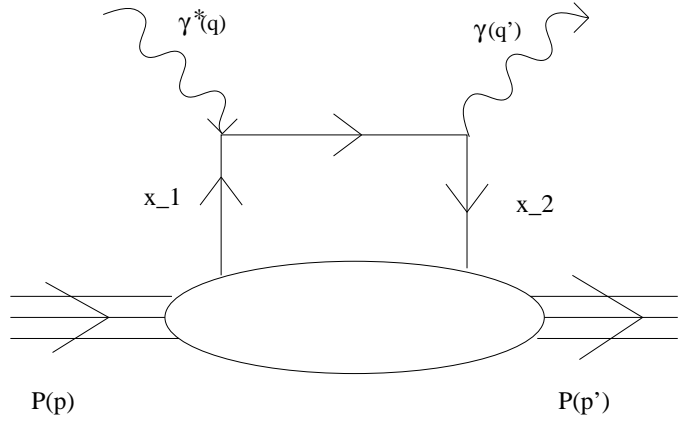,height=5.5cm}}
\caption{The lowest order handbag contribution to DVCS with $Q^2=-q^2$ and 
$q^{\prime 2}=0$.}
\label{hand}
\end{figure}

The important kinematical parameters are the following: $Q^2$
is the virtuality of the probing photon, $t=(p-p')^2$ is the momentum transfer 
to the final state proton, $x_{Bj}$ is the usual Bjorken scaling 
variable, $\Delta=x_1-x_2$ is
 the asymmetry parameter of the process and $x_1$ and $x_2$ 
are the longitudinal light-cone momentum fractions of the partons connecting 
the intermediate state to the hard scattering part (see Fig.\ 1). Note that 
in the case of DVCS $x_{Bj}=\Delta$.

For our study we need the definitions of nondiagonal parton distributions in 
terms of light-cone operators, their evolution equations and LO evolution 
kernels which we will review next.

The definitions of nondiagonal quark and gluon distributions are given by 
\cite{2,6a,8}
\bea
& &f_{q/p}= \int^{\infty}_{-\infty}\frac{dy^-}{4\pi}e^{-ix_2p^+y^-}\langle p|
T\bar \psi(0,y^-,{\bf 0_{\perp}})\gamma^+{\it P}\psi(0)|p'\rangle \ ,\nonumber\\
& &f_{g/p}= -\int^{\infty}_{-\infty}\frac{dy^-}{2\pi}\frac{1}{x_1x_2p^+}
e^{-ix_2p^+y^-}\langle p|
T G_{\nu}^+(0,y^-,{\bf 0_{\perp}}){\it P}G^{\nu+}(0)|p'\rangle \ ,
\eea 
with the following evolution equations \cite{2,4,5,6a}
for the singlet (S) and nonsignlet (NS) case

\bea
& &\frac{dq_{NS}(x_{1},\Delta,Q^2)}{d\ln{Q^2}}=\int^{1}_{x_{1}} 
\frac{dy_{1}}{y_{1}}P^{NS}_{qq}q_{NS}(y_{1},\Delta,Q_0^2) \ ,\nonumber\\
& &\frac{dg_S(x_{1},\Delta,Q^2)}{d\ln{Q^2}}=\int^{1}_{x_{1}} 
\frac{dy_{1}}{y_{1}}\left [ P^{S}_{gg}g_S(y_{1},\Delta,Q_0^2)+P^{S}_{gq}
q_S(y_{1},\Delta,Q_0^2)\right ] \ ,\nonumber\\
& &\frac{dq_S(x_{1},\Delta,Q^2)}{d\ln{Q^2}}=\int^{1}_{x_{1}} 
\frac{dy_{1}}{y_{1}}\left [ P^{S}_{qq}q_S(y_{1},\Delta,Q_0^2)+P^{S}_{qg}
g_S(y_{1},\Delta,Q_0^2)\right ] \ ,
\label{evol}
\eea
and evolution kernels \cite{2,4,5,6a}
\bea
P^{S,NS}_{qq}(x_1/y_1,\Delta/y_1)&=&\frac{\alpha_s}{\pi} C_f [ 
\frac{\frac{x_1}{y_1} + \frac{x_1^3}{y_1^3} - \frac{\Delta}{y_1} 
(\frac{x_1}{y_1} + \frac{x_1^{2}}{y_1^2})}{(1 - \frac{\Delta}{y_1}) 
(1 - \frac{x_1}{y_1})_{+}} +\frac{3}{2}\delta (1-\frac{x_1}{y_1}) \ ,\nonumber\\
P^{S}_{qg} (x_1/y_1, \Delta/y_1) &=& \frac{\alpha_s}{\pi} N_F \frac{
[\frac{x_1^{3}}{y_1^3} + \frac{x_1}{y_1}(1 - \frac{x_1}{y_1})^{2} - 
\frac{x_1^{2}}{y_1^2}\frac{\Delta}{y_1}]}{(1 - \frac{\Delta}{y_1})^2} \ ,
\nonumber\\
P^{S}_{gq}(x_1/y_1, \Delta/y_1) &=& \frac{\alpha_s}{\pi} C_F \frac{[ 1 + 
(1-\frac{x_1}{y_1})^{2} - \frac{\Delta}{y_1}]}{1 - \frac{\Delta}{y_1}} \ ,
\nonumber\\
P^{S}_{gg}(x_1/y_1, \Delta/y_1)&=&N_{c} [2\frac{(1 - \frac{x_1}{y_1})^2 + 
(\frac{1}{2} - \frac{x_1^{2}}{y_1^2})(\frac{x_1 - \Delta}{y_1})}
{(1 - \frac{\Delta}{y_1})^2} - 1 - \frac{x_1}{y_1} + 
\frac{1}{(1 - \frac{x_1}{y_1})}_{+}\nonumber\\
& & + \frac{\frac{x_1}{y_1} - \frac{\Delta}{y_1}}
{(1 - \frac{x_1}{y_1})(1 - \frac{\Delta}{y_1})}_+ + \delta (1-\frac{x_1}{y_1}) 
\left [ \frac{\beta_0}{2N_C} \right ]] \ ,
\label{kern}
\eea
with the generalized $+$ regularization prescription necessary for the 
nondiagonal case which is described in detail in Ref.\ \cite{2,5,6a}.

\section{Radyushkin's model}
\label{sec:mod}

In the following we will briefly review the simplest model for an initial 
nondiagonal parton distribution as proposed by Radyushkin in Ref.\ \cite{20}.
The starting point are double distributions 
$F(x,y) = F(x,y,t=0)$ which behave like a distribution amplitude in the 
variable $y$ and like a parton distribution in the variable $x$. They are 
expressed through multivariable Fourier transforms of matrix elements of 
bilocal, renormalized, light-cone operators \cite{2,20}. Symmetry and
spectral conditions suggest the following ansatz for the double distribution 
\cite{20}
\beq
F(x,y) = \frac{h(x,y)}{h(x)}f(x)
\label{ansatz}
\eeq
such that with the normalization condition
\beq
h(x) = \int_{0}^{1-x} h(x,y) dy \ ,
\eeq
one obtains the diagonal distribution $f(x)$
\beq
f(x) = \int_{0}^{1-x}F(x,y)dy \ .
\eeq 

The simplest realization of the ansatz Eq.\ (\ref{ansatz}) is \cite{20}
\beq
F(x,y,Q_0) = \delta (y-k(1-x))f(x) \ .
\label{specans}
\eeq
 The $\delta$-function in $y$ prevents any spread of the double 
distribution in the $y$ direction and $k$ is
 a number smaller than $1$ and is determined by the powerlaw behavior 
of parton densities at small $x$.   

In order to obtain the nondiagonal parton distribution in the DGLAP region 
($x_1>\Delta$) from the double distribution, one has to integrate over $y$ with
an explicit $\Delta$ dependence
\beq
F(x_1,\Delta,Q_0)_i = \int_{0}^{\frac{1-x_1}{1-\Delta}} F_i(x_1-\Delta y,y,Q_0)
dy \ .
\eeq 
Using Eq.\ (\ref{specans}) one finds the following initial distribution which 
will be our input for the evolution 
\beq
F_{i}(x_1,\Delta,Q_0) = \frac{1}{1-k\Delta}f_{i}\left(\frac{x_1-k\Delta}
{1-k\Delta},Q_0\right ) \ ,
\label{ans1}
\eeq
with $i=q, \bar q, g$ and we chose $k=0.5$ which is in line with the arguments 
in \cite{20} that the crest of the double distribution is shifted towards the 
line $y=(1-x)/2$.
Note that the ansatz of the previous numerical studies \cite{5,7}
was that the 
diagonal and nondiagonal distributions were equal in the normalization point 
which corresponds to a double distribution similar to Eq.\ (\ref{specans}) 
with $\delta (y-k(1-x))\rightarrow \delta (y)$.

The above made ansatz leads to the following predictions (see Ref.\ \cite{20}
for more details) for nondiagonal parton distributions:
\beq
R(Q) = \frac{F_g(\Delta,\Delta,Q)}{\Delta f_g(\Delta,Q)} \simeq 
\frac{(\Delta/2)f_g(\Delta/2,Q)}{\Delta f_g(\Delta,Q)}
\label{pred1}
\eeq
and
\beq
F_{i}(x_1,\Delta,Q) \simeq (x_1-\Delta/2)f_{i}(x-\Delta/2,Q) \ .
\label{pred2}
\eeq
The first equation is of importance in the case of DVCS since there one has 
$x_1 = \Delta = x_{Bj}$. The second equation supposedly holds for any $Q$. If 
the above predictions hold true even after evolution it would give a very 
useful approximation of nondiagonal parton densities.

\section{Results of Evolution}
\label{sec:res}

Our input distributions for the diagonal parton densities
 $f(x)$ in Eq.\ (\ref{specans}) will be CTEQ4M and CTEQ4LQ \cite{21}. 
The reason why we chose these distributions is simple.
Both CTEQ4M at a fairly high normalization point of
$Q_0=1.6~\mbox{GeV}$ and CTEQ4LQ at a very low normalization point of 
$Q_0 = 0.7~\mbox{GeV}$ are discernably different in the initial shape. 
This  will help us discriminate fairly easily 
how robust to the initial shape the predictions of the previous section are.

Our input distributions for the nondiagonal evolution are those satisfying  
relation\ (\ref{ans1}). Since the nondiagonal evolution kernels are known 
explicitly to leading order (LO) only, we only use LO kernels for the diagonal
evolution also, in order to have a consistent comparison between the nondiagonal 
and diagonal case. 

Let us first discuss how well Eq.\ (\ref{pred2}) is satisfied.
The results of our numerical study are given in 
Figs.\ (\ref{fig4},\ref{fig5},\ref{fig6},\ref{fig7}). Although the input 
distribution for the nondiagonal evolution was chosen according to 
Eq.\ (\ref{pred2}) at $Q=Q_{0}$, the evolution  hardly violates this relationship 
at 
$Q > Q_{0}$. One can see from Figs.\ (\ref{fig4},\ref{fig5}) that as the 
nondiagonal density evolves with $Q$ the ratio of the nondiagonal to diagonal 
gluon parton density 
\mbox{$g(x_{1},\Delta,Q)/((x_{1}-\Delta/2)G(x_{1}-\Delta/2,Q))$} stays within 
$10\%$ for all $x_{1}$ and $Q$, for both CTEQ4M and CTEQ4LQ.
Note that in these figures 
$x_{1 min}=1.0001 \Delta$. As $x_{1}$ increases the difference between the 
nondiagonal and diagonal densities becomes small. This is a natural behaviour 
of the nondiagonal densities since for $x_{1} \gg \Delta$ all the asymmetry 
related effects are unimportant.

Next, we present the ratio of the nondiagonal to diagonal quark parton 
densities\\
\mbox{$q(x_{1},\Delta,Q)/((x_{1}-\Delta/2)Q(x_{1}-\Delta/2,Q))$}
in Figs.\ (\ref{fig6},\ref{fig7}). The quark parton density is defined
\begin{equation}
Q(x,Q)=u(x,Q)+d(x,Q)+s(x,Q)+\bar{u}(x,Q)+\bar{d}(x,Q)+\bar{s}(x,Q) \ .
\end{equation}
Here we observe a similar tendency as in the case of gluons -- 
at $x_{1}$ close
to $\Delta$ -- the difference between the nondiagonal and diagonal densities is
different from $1$. 
However, in the case of quarks the ratio is significantly larger than $1$. In 
fact, the nondiagonal quark is about $3$ times larger than the diagonal one
at $x_{1} \simeq \Delta$  (here $x_{1 min}=1.0001 \Delta$)
\footnote{The same behaviour was observed by Golec-Biernat \cite{22}}.
This result is not too surprising, in light of the findings of Ref.\ \cite{6a},
which showed a large deviation of the nondiagonal quark distribution from the 
diagonal one for $x_1\simeq \Delta$ with a strong enhancement in the deviation 
for a low normalization point.
As $x_{1}$ becomes significantly large than $\Delta$ the ratio quickly and
smoothly approaches $1$, as expected.

To summarize this set of figures, we conclude that the prescription 
of Eq.\ (\ref{pred2}), where one shifts the argument of the diagonal parton 
density by $\Delta/2$, decreases the percentage deviation of the nondiagonal 
to diagonal parton density by approximately a factor of $3-4$ for gluons  
(compare to the relevant figures from Ref.\ \cite{6a}), giving a very good 
agreement for all $x$ and $Q$. For quarks the approximation of 
Eq.\ (\ref{pred2}) is much worse as compared to the gluon case, however it becomes
relatively good for $x_1 \geq 2$ to $3\Delta$.

\begin{figure}
\centering
\vskip -3cm
\mbox{\epsfig{file=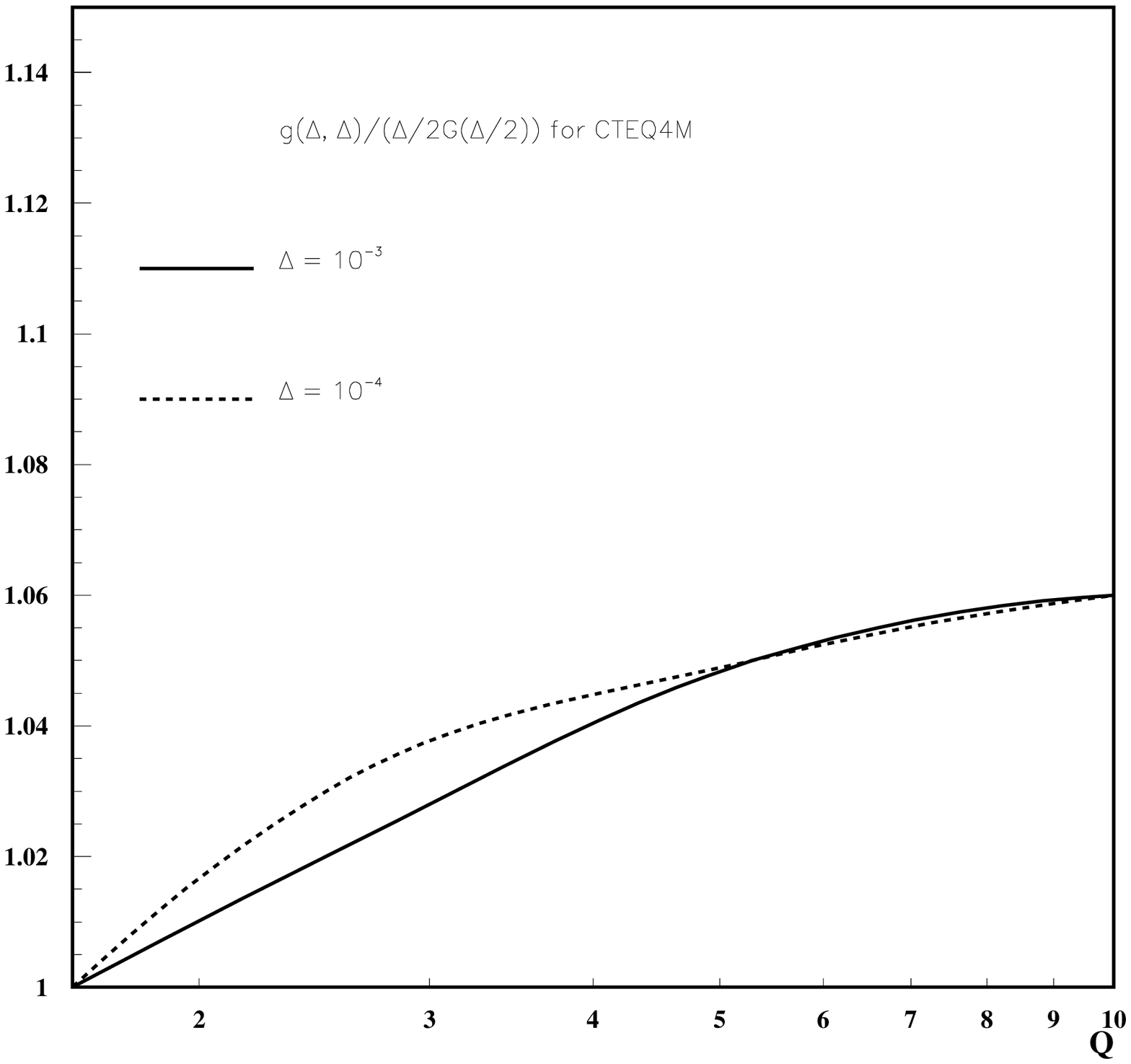,height=15cm}}
\vskip -3cm
\mbox{\epsfig{file=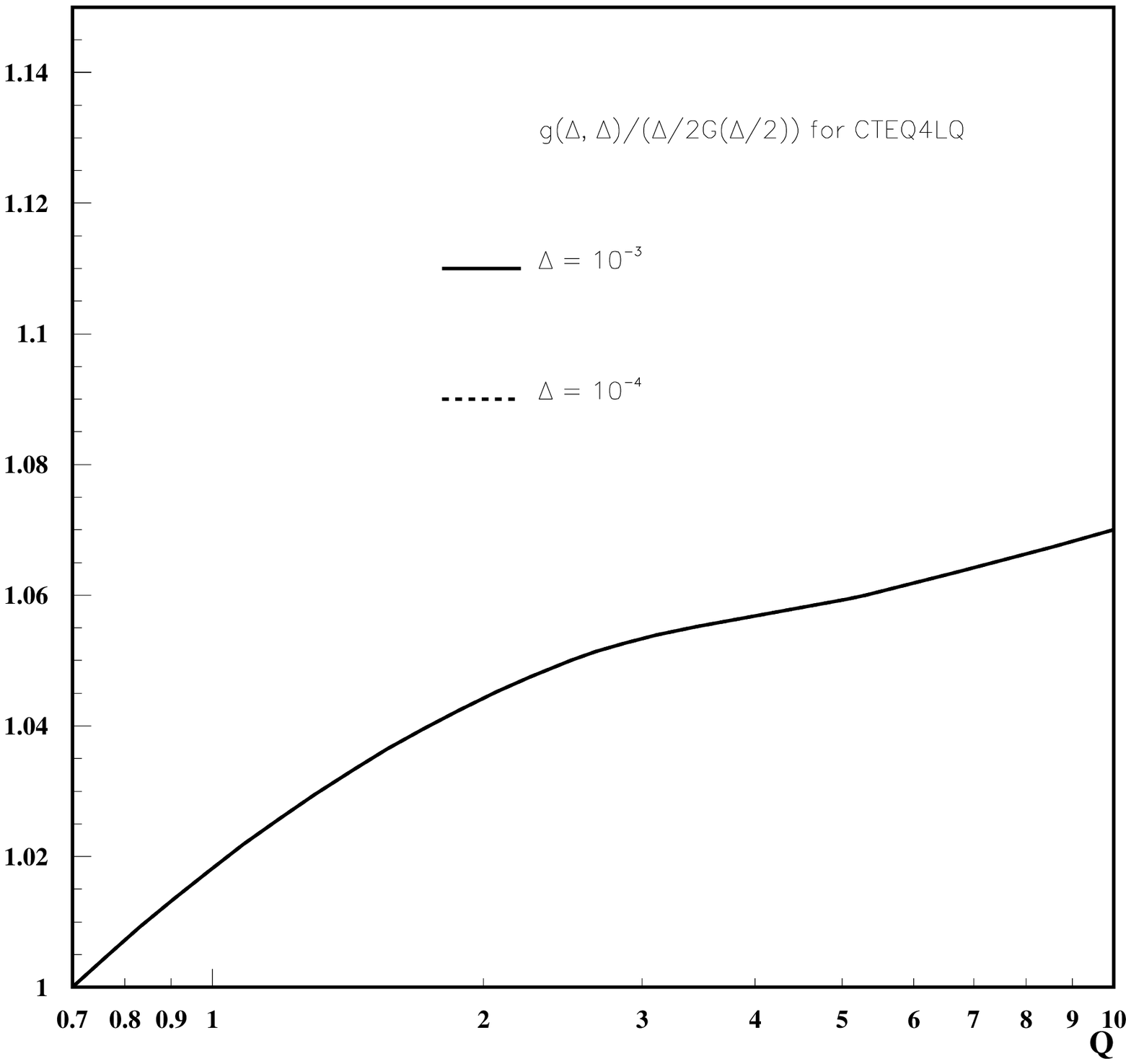,height=15cm}}
\vskip -1.5cm
\caption{$R(Q)$ is ploted as a function of $Q$ at $\Delta=10^{-4}$ 
and $\Delta=10^{-3}$. The input distributions are CTEQ4M and CTEQ4LQ.}
\label{fig2}  
\end{figure}

\begin{figure}
\centering
\vskip -3cm
\mbox{\epsfig{file=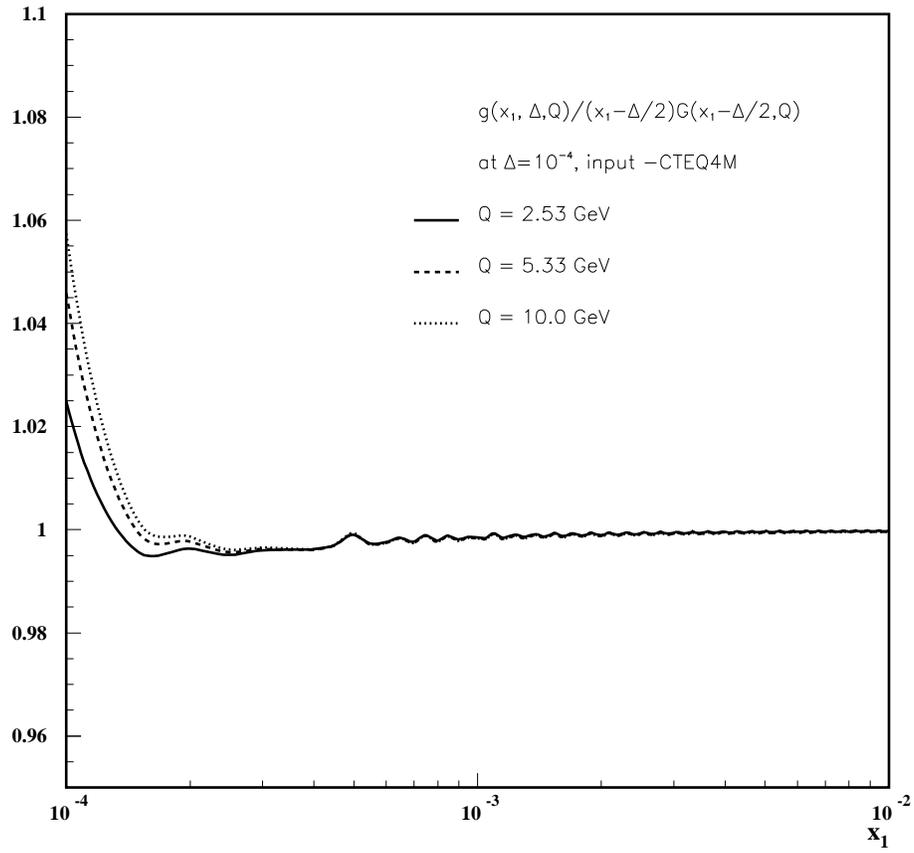,height=15cm}}
\vskip -3.5cm
\mbox{\epsfig{file=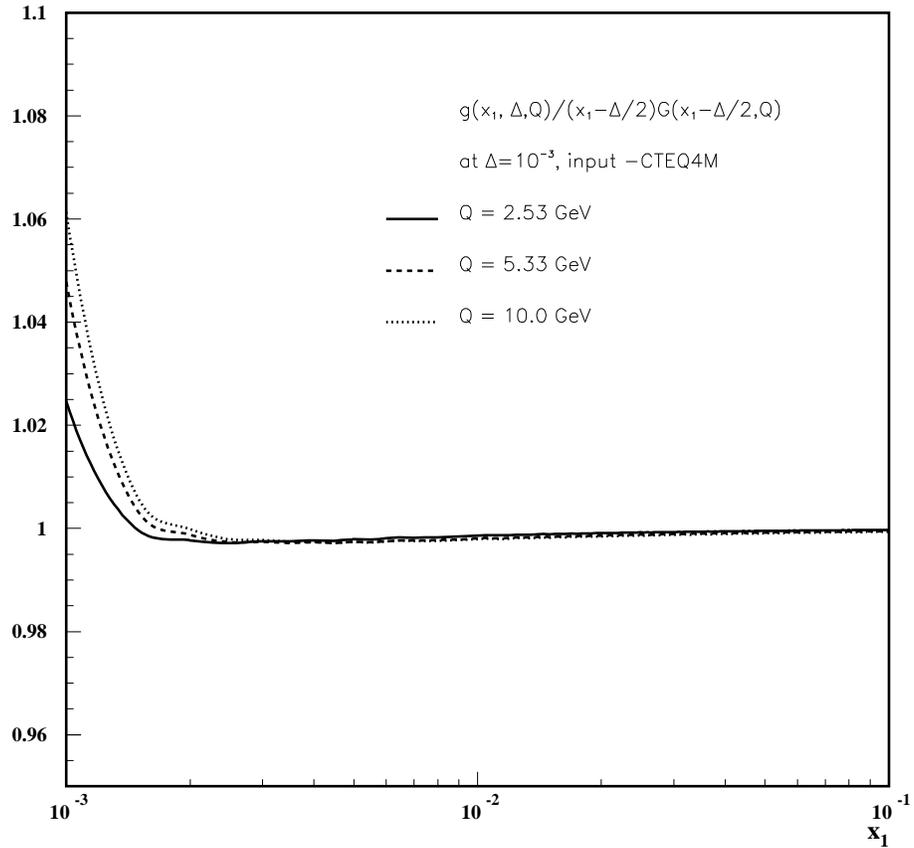,height=15cm}}
\vskip -2cm
\caption{
\mbox{$g(x_{1},\Delta,Q)/(x_{1}-\Delta/2)G(x_{1}-\Delta/2,Q)$}
 as a function of $x_{1}$ for $Q$=1.6, 2.5, 5.3, 10.0 GeV for $\Delta=10^{-4}$ 
 and $\Delta=10^{-3}$. The initial distribution is CTEQ4M.}  
\label{fig4}
\end{figure}

\begin{figure}
\centering
\vskip -3cm
\mbox{\epsfig{file=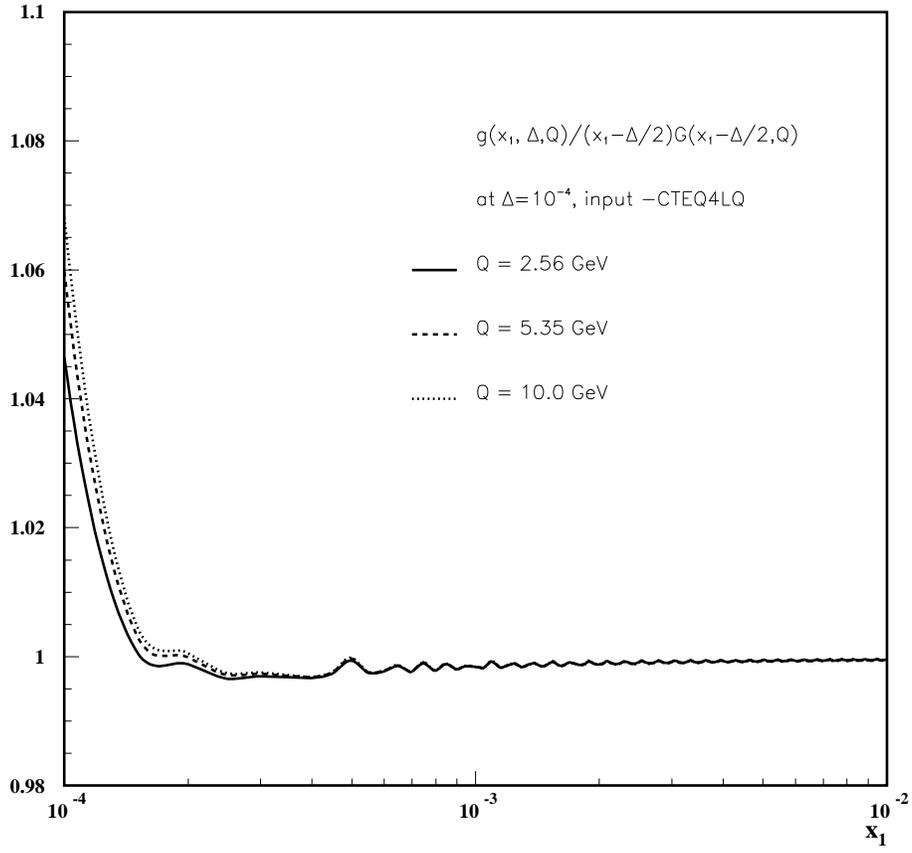,height=15cm}}
\vskip -3.5cm
\mbox{\epsfig{file=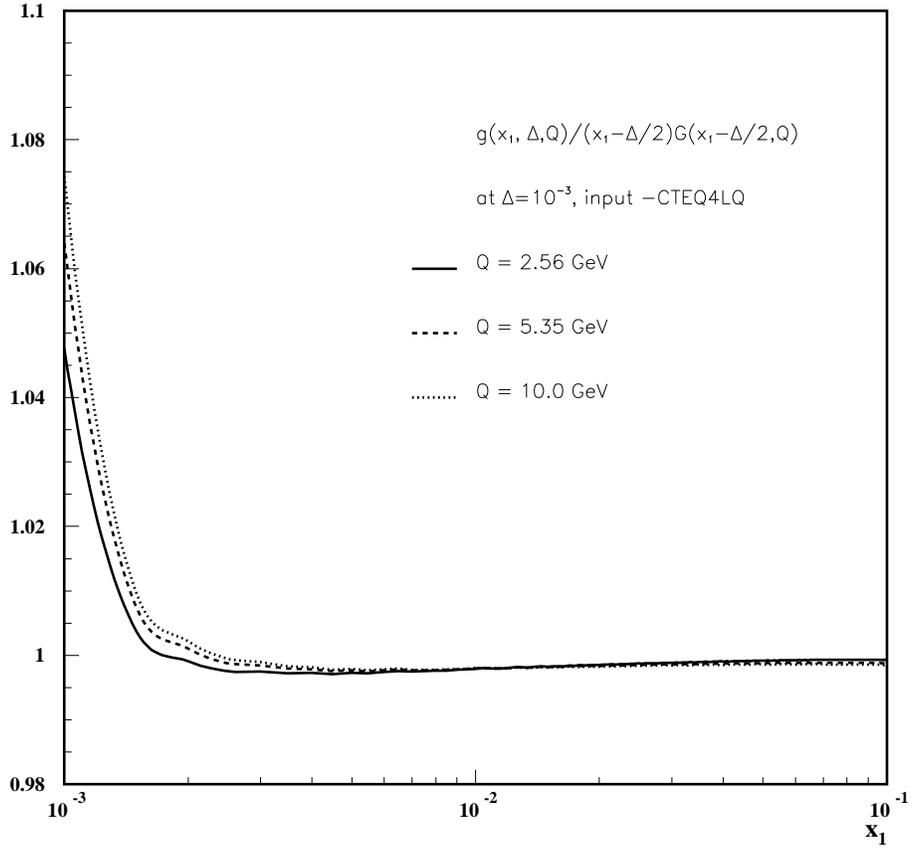,height=15cm}}
\vskip -2cm
\caption{
\mbox{$g(x_{1},\Delta,Q) / (x_{1}-\Delta /2)G(x_{1}-\Delta /2,Q)$}
 as a function of $x_{1}$ for $Q$=0.7, 2.8, 5.9, 10.0 GeV for $\Delta=10^{-4}$ 
 and $\Delta=10^{-3}$. The initial distribution is CTEQ4LQ.}  
\label{fig5}
\end{figure}

\begin{figure}
\centering
\vskip -3cm
\mbox{\epsfig{file=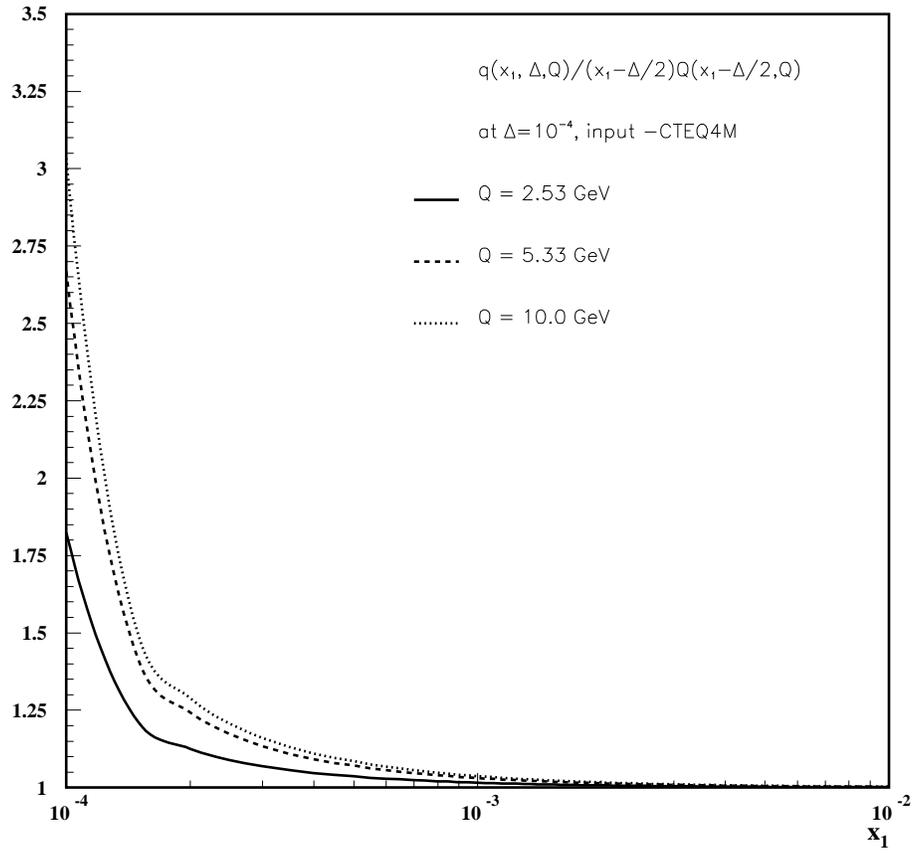,height=15cm}}
\vskip -3.5cm
\mbox{\epsfig{file=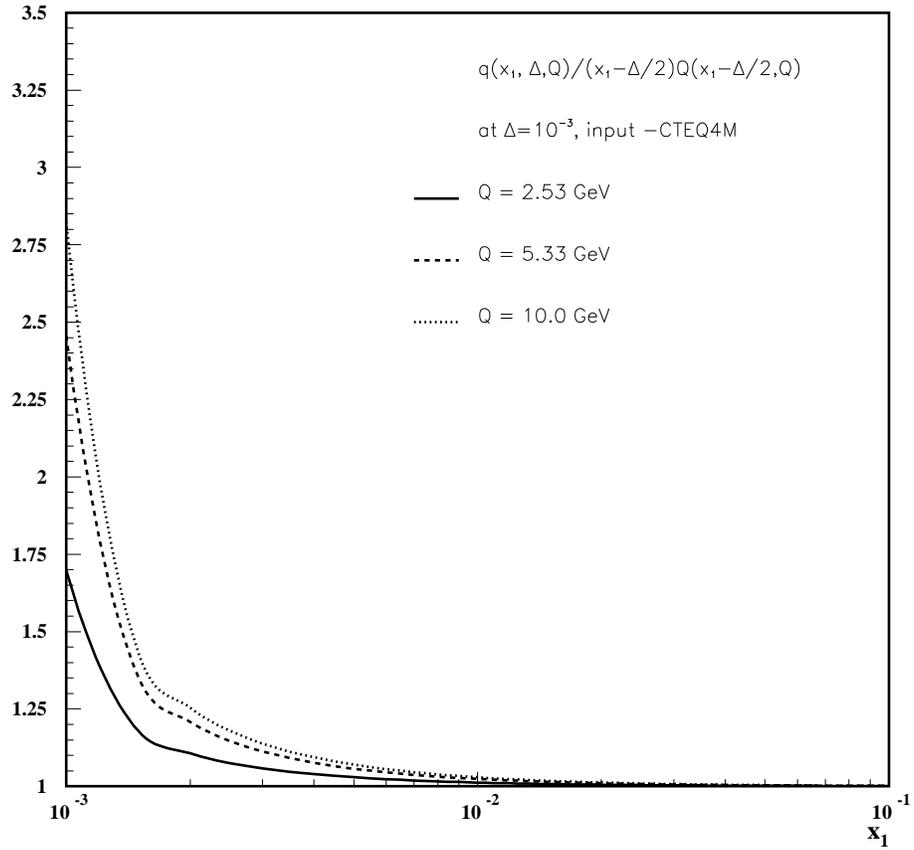,height=15cm}}
\vskip -2cm
\caption{
\mbox{$q(x_{1},\Delta,Q) / (x_{1}-\Delta /2)Q(x_{1}-\Delta /2,Q)$}
 as a function of $x_{1}$ for $Q$=1.6, 2.5, 5.3, 10.0 GeV for $\Delta=10^{-4}$ 
 and $\Delta=10^{-3}$. The initial distribution is CTEQ4M.}  
\label{fig6}
\end{figure}

\begin{figure}
\centering
\vskip -3cm
\mbox{\epsfig{file=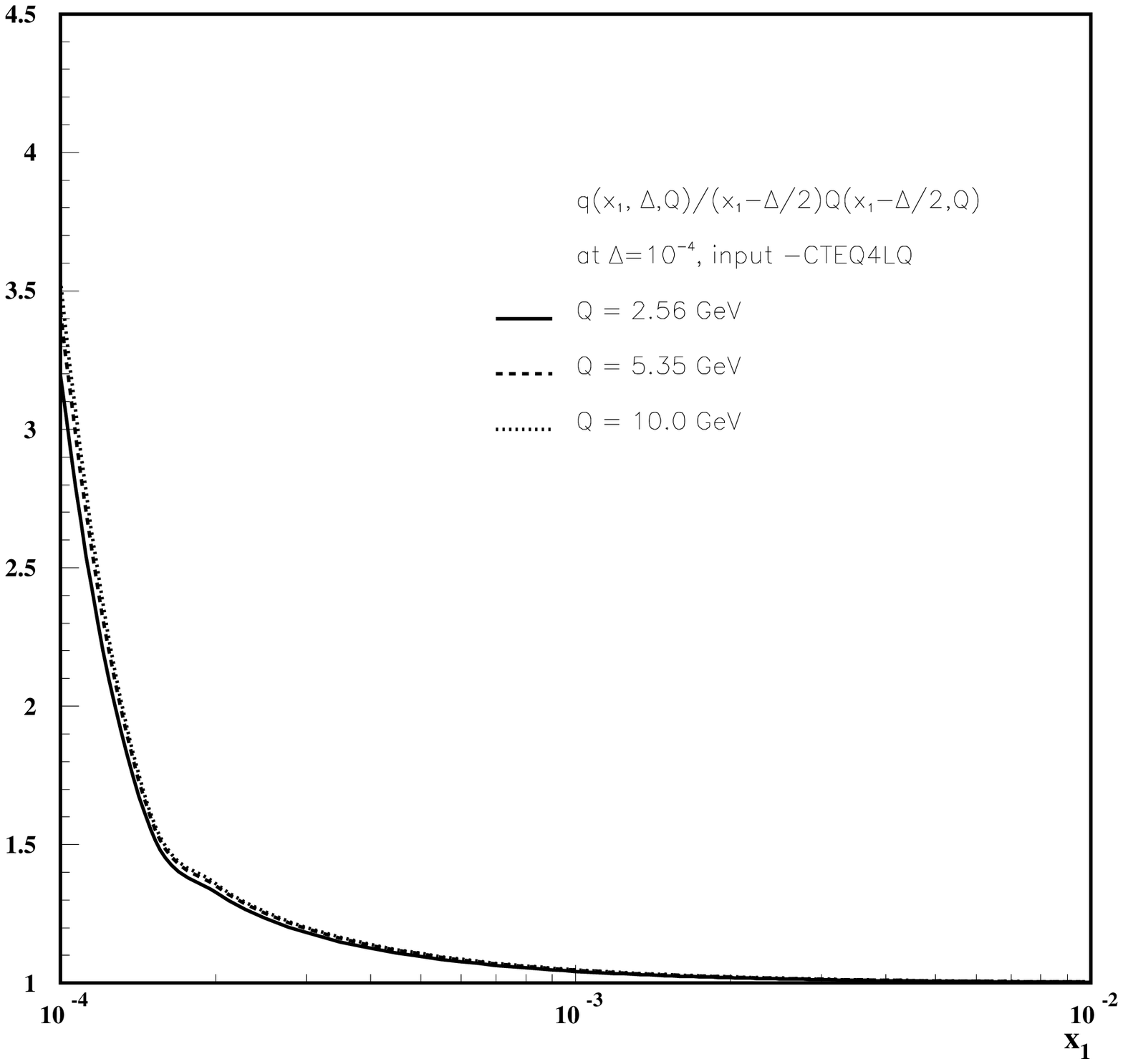,height=15cm}}
\vskip -3.5cm
\mbox{\epsfig{file=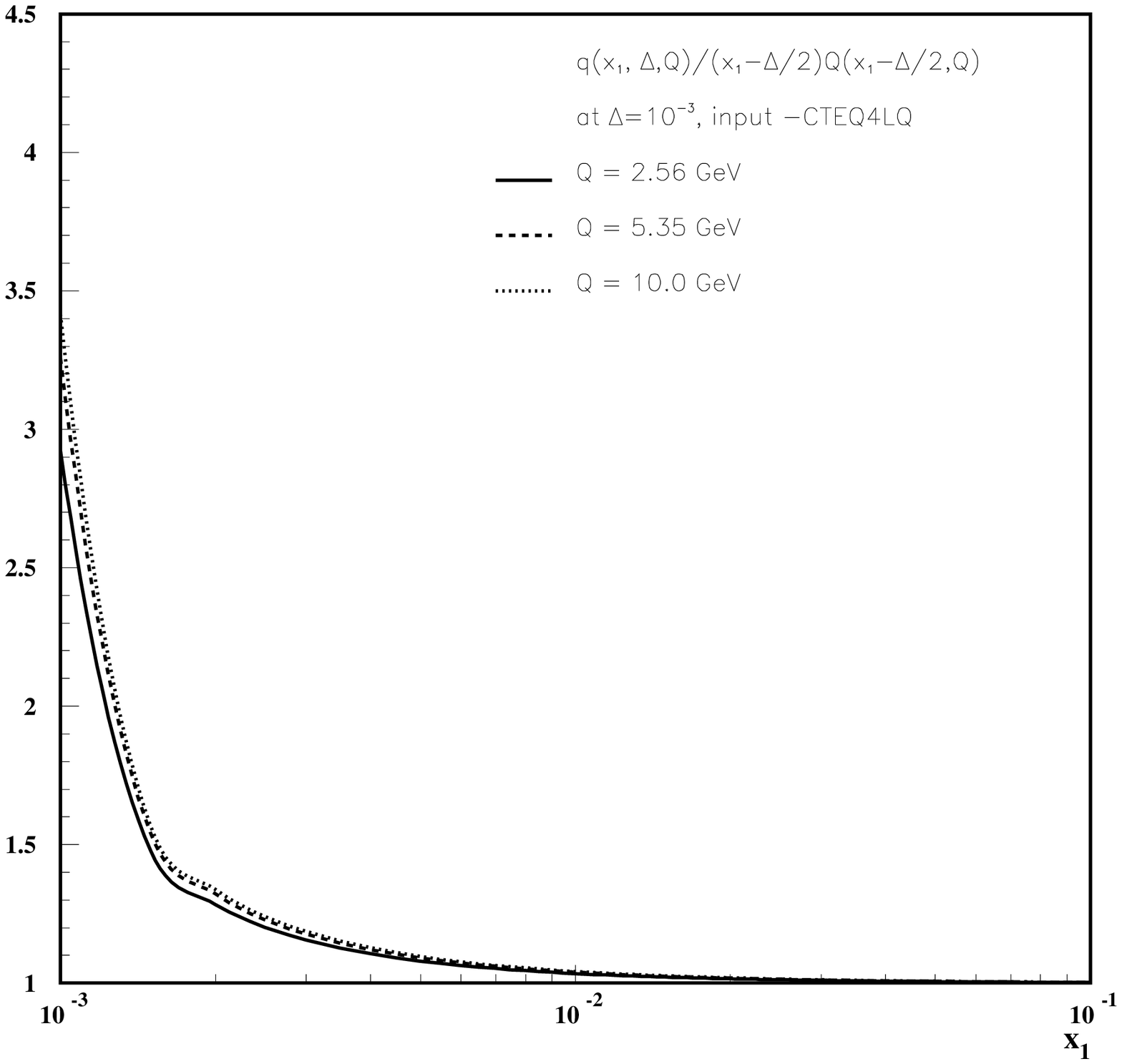,height=15cm}}
\vskip -2cm
\caption{
\mbox{$q(x_{1},\Delta,Q)/(x_{1}-\Delta /2)Q(x_{1}-\Delta /2,Q)$}
 as a function of $x_{1}$ for $Q$=0.7, 2.8, 5.9, 10.0 GeV for $\Delta=10^{-4}$ 
 and $\Delta=10^{-3}$. The initial distribution is CTEQ4LQ.}  
\label{fig7}
\end{figure}

Next we find that Eq.\ (\ref{pred1}) is fulfilled with an accuracy better than
$8\%$ in the $x$ and $Q$ range studied for both CTEQ4M and CTEQ4LQ
(see Fig.\ \ref{fig2}). This is very good as far as DVCS studies are concerned
since Eq.\ (\ref{pred1}) is an all order statement and thus one can use the NLO evolution
of diagonal gluon densities to make NLO predictions for DVCS!
As mentioned before, we have chosen $x_{2}=x_{1}-\Delta=10^{-8}$ for $x_{1}=10^{-4}$ and 
$x_{2}=10^{-7}$ for $x_{1}=10^{-3}$. As one can see there is no big difference
in the ratio of the densities for the two $\Delta$ values studied which is in agreement with 
Radyushkin's statement that Eq.\ (\ref{pred1}) should hold as long as $\Delta$
is much smaller than $1$.

Finally, we would like, in light of the previous findings, comment on the
NLO evolution in the DGLAP region. Given the fact that 
Eq.\ (\ref{pred1},\ref{pred2}) was based on general arguments in \cite{20} 
and that it holds in LO evolution with a certain degree of accuracy, 
we predict that the NLO evolution will not 
change the above relations, in other words, that the NLO evolution of the 
nondiagonal gluon distribution can be predicted to
a similar accuracy by the NLO evolution of the diagonal gluon distribution. 
We base this statement on the results of the above analysis and the 
observations of 
\cite{6a} that the NLO  corrections of the nondiagonal evolution should be in 
the same direction as in the diagonal case, which reduces the LO results, and 
of the same magnitude. 
The former statement is due to the observations made in \cite{6a} that if, in
the nondiagonal case, the NLO corrections were in the opposite direction,
which would lead to a 
marked deviation from the LO results, compared to the diagonal
case, the overall sign of the NLO nondiagonal kernels would have to change for
some $\Delta \neq 0$ since in the limit $\Delta \rightarrow 0$ we have to 
recover the diagonal case. This occurance is not likely for the following 
reason: First, the Feynaman diagrams involved in the calculation of the NLO 
nondiagonal kernels are the same as in the diagonal case, except for the 
different kinematics, therefore, we have a very good idea about the type 
of terms appearing in the kernels, namely polynomials, logs and terms in need 
of regularization such as $\ln(z) \times \frac{\ln (1-z)}{(1-z)}$. Secondly, the 
kernels, as 
stated before, have to reduce to the diagonal case in the limit of vanishing 
$\Delta$ which fixes the sign of most terms in the kernel, thus the only type
of terms which are allowed and could change the overall sign of the kernel are
of the form 
\beq
\frac{\Delta}{y_1} f(x_1/y_1,\Delta/y_1) 
\label{term}
\eeq
which will be numerically 
small unless $y_1 \simeq \Delta$ in the convolution integral of the 
evolution equations. Moreover, we know that in this limit the contribution of 
the regularized terms in the kernel gives the largest contributions in the 
convolution integral and therefore sign changing contributions in the 
nondiagonal case would have to originate from regularized terms. This in turn
disallows a term like Eq.\ \ref{term} due to the fact that regularized terms
are not allowed to vanish in the diagonal limit, since the regularized 
 terms arise from the same Feynman diagrams in the both  
diagonal and nondiagonal case. Therefore, the overall sign of the contribution
of the NLO nondiagonal kernels will be of the same as in the diagonal case
. In addition, the magnitude of the correction should be of the same magnitude 
as in the diagonal case since one has the conditions $R(Q)>1$ \cite{6,18,20}, 
$R(Q)<1.5$ \cite{20} for the gluon distribution and the fact that the LO results 
at high $Q$ are already fairly close to the upper bound. This forces the NLO 
corrections in the nondiagonal case not to exceed the diagonal corrections by 
a factor of $3$ or so, lest it violates the boundary conditions for $R(Q)$.

\section{Conclusions}
\label{sec:concl} 

In the above, we examined predictions made in Ref.\ \cite{20} about 
relationships between nondiagonal and diagonal parton distributions in the 
DGLAP region based on certain models for nondiagonal parton distributions in
the normalization point. We found that the evolution does not destroy the 
validity of Eqs.\ (\ref{pred1},\ref{pred2}) for all $x$ and $Q > Q_{0}$ for
both CTEQ4M and CTEQ4LQ in the case of gluons and for $x_1 \geq 2 \div 3\Delta$
for quarks. Therefore, we conclude that Eqs.\ (\ref{pred1},\ref{pred2}) do 
supply a reliable approximation of nondiagonal parton densities for 
$\Delta \ll 1$.
We also conclude that we can hardly see a variance of the accuracy of predictions\ 
(\ref{pred1},\ref{pred2}) for different initial distributions. The accuracy is 
slightly better for the one which supplies a less steep gluon density at small $x$.
 
Based on these results and the results from
Ref.\ \cite{6a}, we predicted the NLO evolution of the nondiagonal gluon 
distribution to be within $\approx 20\%$ of the diagonal gluon distribution 
for the above made ansatz and for a large range of $Q$.

\section*{Acknowledgments}

This work was supported in part by the U.S.\ Department of Energy
under grant number DE-FG02-90ER-40577.
We would like to thank John Collins, Mark Strikman and Anatoly Radyushkin for 
helpful conversations and once more  Anatoly Radyushkin for pointing out
an inconsistency in an earlier draft version due to a minor error in our 
evolution code.  

We also would like to thank Martin McDermott who has drawn our attention to an 
error in the input of the initial quark parton densities in our evolution code.
This forced us to review the numerical results and figures of this paper and 
our previous one \cite{6a}.

After this work was complete, we learned that M\"uller et al.\ \cite{23} 
had performed a numerical study of the NLO effects for the non-singlet and 
singlet distributions and found that the corrections were within $10-30\%$ 
percent of the LO result confirming our statements on the NLO corrections.

\newpage

\newpage

\newpage

\newpage

\newpage

\begin{references}

\bibitem{1}     S.J. Brodsky, L.L. Frankfurt, J.F. Gunion, A.H. Mueller, 
                    and M. Strikman, Phys. Rev. {\bf D50} (1994) 3134; 
                    {\it ibid.} Erratum in print

\bibitem{2}     A. Radyushkin Phys. Letters {\bf B385} (1996) 333,
                   Phys.Lett {\bf B380} (1996) 417, Phys.\ Rev.\ 
	           {\bf D56}, 5524 (1997).

\bibitem{3}     J.C. Collins, L. Frankfurt, and M. Strikman, 
                    Phys. Rev. {\bf D56} (1997) 2982.

\bibitem{4}     X.-D. Ji, Phys. Rev. {\bf D55} (1997) 7114, Phys.\ Rev.\
                Lett.\ {\bf 78}, 610 (1997).

\bibitem{5}     L.L. Frankfurt, A. Freund, V. Guzey and M. Strikman,
                Phys.\ Lett.\ {\bf B 418}, 345 (1998).

\bibitem{6}     A.Martin and M.Ryskin, Phys.\ Rev.\ {\bf D57}, 6692 (1998).

\bibitem{6a}    A. Freund and V. Guzey, hep-ph/9801388.

\bibitem{7}     L. Mankiewicz, G. Piller and T. Weigel, hep-ph/9711227.

\bibitem{8}     J.C. Collins and A. Freund, hep-ph/9801262.

\bibitem{9}     X.-D. Ji and J. Osborne, hep-ph/9801260.

\bibitem{10}    D. M\"uller, ``Restricted conformal invariance in QCD and its
                predictive power for virtual photon processes'',
                hep-ph/9704406.

\bibitem{11}    X. Ji and J. Osborne, Phys.\ Rev.\ {\bf D57}, 1337  (1998).

\bibitem{12}    A.V. Belitsky and D. M\"uller, Phys.\ Lett.\ {\bf B417}, 129
                (1998), hep-ph/9802411, hep-ph/9804379.

\bibitem{12a}   A.V. Belitsky, B. Geyer, D. M\"uller, A. Sch\"afer
                Phys.\ Lett.\ {bf B421}, 312 (1998).

\bibitem{13}    M. Diehl, T. Gousset, B. Pire, and J.P. Ralston,
                Phys.\ Lett.\ {\bf B411}, 193 (1997),
   
\bibitem{14}    Z. Chen, ``Non-Forward and Unequal Mass Virtual Compton
                Scattering'' , hep-ph/9705279.

\bibitem{15}    L. Frankfurt, A. Freund and M. Strikman, hep-ph/9710356 to 
                appear in Phys.\ Rev.\ D.

\bibitem{16}    L. Mankiewicz, G. Piller, E. Stein, M. V\"attinen and T. Weigl,
                ``NLO Corrections to Deeply-Virtual Compton Scattering'',
                hep-ph/9712251.

\bibitem{17}    J. Bl\"umlein, B. Geyer, and D. Robaschik, Phys.\ Lett.\ 
                {\bf B406}, 161 (1997), hep-ph/9711405.
        
\bibitem{18}    B. Pire, J. Soffer and O. V. Teryaev, hep-ph/9804284.

\bibitem{19}    L. Frankfurt, A. Freund and M. Strikman preprint in 
                preparation. 

\bibitem{11a}   F.\ M.\ Dittes, J.\ Horejsi, B.\ Geyer, D.\ M\"uller and
                D.\ Robaschick, Phys.\ Lett.\ {\bf B209}, 325 (1988). 

\bibitem{20}    A. Radyushkin, hep-ph/9805342.

\bibitem{21}    H. Lai {\it et al.}, Phys. Rev. {\bf D55}, 1280 (1997).

\bibitem{22}    K. Golec-Biernat, private communications.

\bibitem{23}    A.\ V.\ Belitsky, D.\ M\"uller, L.\ Niedermeier and A.\
                Sch\"afer, hep-ph/9806232 and hep-ph/9810275.


\end{references}
\end{document}